\documentstyle[epsf,12pt]{article}
\newcommand{\lsim}{\mbox{\raisebox{-.3em}{$\stackrel{<}{\sim}$}}}
\renewcommand{\thefootnote}{\fnsymbol{footnote}}

\renewcommand{\cite}[1]{\ref{#1}}

\newcommand{\half}{\frac{1}{2}}
\newcommand{\reflef}{(\ref}
\newcommand{\beq}{\begin{equation}}
\newcommand{\eeq}{\end{equation}}
\newcommand{\beqa}{\begin{eqnarray}}
\newcommand{\eeqa}{\end{eqnarray}}
\newcommand{\bcent}{\begin{center}}
\newcommand{\ecent}{\end{center}}

\begin{document}


\baselineskip=0.8cm
\bcent
{\Large\bf A two-scalar model for a small but nonzero cosmological constant}\\[2.em]
\baselineskip=0.6cm
Yasunori Fujii\footnote{E-mail address: fujii@handy.n-fukushi.ac.jp}\vspace{.6em}\\
Nihon Fukushi University, Handa, 475-0012\ Japan\\
\ecent

\mbox{}\\[0.0em]
\bcent
{\large\bf Abstract}\\[1.em]
\baselineskip=0.5cm
\parbox{13cm}{
We revisit a model of the two-scalar system proposed previously for
understanding a small but nonzero cosmological constant.  The model
provides solutions of the scalar-fields energy $\rho_s$ which behaves
truly constant for a limited time interval rather than in the way of
tracker- or scaling-type variations.  This causes a mini-inflation, as 
indicated by recent observations.  As another novel feature, $\rho_s$ 
and the ordinary matter density $\rho_m$ fall off always side by side, 
but interlacing, also like (time)$^{-2}$ as an overall behavior in 
conformity with the scenario of a decaying cosmological constant.  A 
mini-inflation occurs whenever $\rho_s$ overtakes $\rho_m$, which may 
happen more than once, shedding a new light on the coincidence problem.  
We present a new example of the solution, and offer an intuitive 
interpretation of the mechanism of the nonlinear dynamics.  We also 
discuss a chaos-like nature of the solution.
}
\ecent


\renewcommand{\thefootnote}{\arabic{footnote}}

\setcounter{footnote}{0}
\baselineskip=0.6cm

\section{Introduction}

Sometime ago we proposed a theoretical model [\cite{plapp}] for the
system of two scalar fields as an effective way to understand a small
but nonzero cosmological constant, as indicated by a number of
observations.  The indication is now even stronger in particular due
to the recent results on type Ia supernovae at high redshift
[\cite{garna},\cite{perlm},\cite{free}].  In accordance with this
development we provide in this paper, (i) an example of the
cosmological solution suited for the recent data, (ii) an intuitive
interpretation of the underlying mechanism of the model, (iii) a novel
way of viewing the ``cosmological coincidence problem."  We also add
comments on the related issues including a generalization, chaos-like
nature of the solution and  choice of a physical conformal frame.

It seems useful to outline beforehand how we arrived at this
particular model, which was conceived in two steps.  We first started
with the prototype Brans-Dicke (BD) model with a constant $\Lambda$ added [\cite{yf3}].  We apply a conformal transformation (Weyl rescaling) to remove the non-minimal coupling term.  We say we have moved from a conformal frame (CF) called J frame after Jordan to  another called E frame after Einstein (denoted by $*$ hereafter if necessary).  In the latter CF we have an exponential potential 

\beq
V(\sigma)=\Lambda e^{-4\zeta\sigma},
\label{ts-1}
\eeq
where $\sigma$ is a canonical scalar field related to the original field $\phi$ by

\beq
\phi =\xi^{-1/2}e^{\zeta\sigma}.
\label{ts-2}
\eeq
The constant $\xi$ is related to the original definition $\omega$ by $4\xi\omega=\epsilon =\pm 1$, and 

\beq
\zeta^{-2} = 6 +\epsilon \xi^{-1},  
\label{ts-3}
\eeq
which should be chosen to be positive so that $\sigma$ is a (non-ghost) normal field.

This exponential potential serves to implement the scenario of a ``decaying cosmological constant," $\Lambda_{\rm eff}=\rho_{\sigma}\sim t_*^{-2}$ (with $t_*$ the cosmic time in E frame), allowing us to understand why today's $\Lambda$ is smaller than the theoretically natural value by as much as 120 orders of magnitude, where $\rho_{\sigma}$ is the energy density of $\sigma$ [\cite{yf1}].

We find, however, that $\rho_{\sigma}$ which falls off in the same way as the ordinary matter density $\rho_{*m}$ is not qualified to explain what the recent observations indicate.  We need $\rho_{\sigma}$ falling off more slowly than $t_*^{-2}$, or more preferably staying constant, at least for some duration of time.  A scalar field expected to show this feature, now often called quintessence, has been a focus of many studies [\cite{qta}-\cite{qtd}].

In this connection we notice that the (analytic) solution resulting in $\rho_{\sigma}\sim t_*^{-2}$ is an attractor realized asymptotically.  We know, on the other hand, that there is an interesting transient solution, which allows $\sigma$ to stay nearly constant temporarily.  This is a consequence of a fast decreasing potential and the frictional force supplied by the cosmological expansion [\cite{fon},\cite{yf3},\cite{qtc}].   The plateau behavior of $\sigma$ lasts until $\rho_{*m}$ comes down to be comparable with the overdamped $\rho_{\sigma}$, when the system goes into the asymptotic state.

It may appear that this plateau mimics a cosmological constant.  It
does, but not to the extent required to fit the observations, because
$\sigma$ is not persevering enough to keep staying constant as it is
seduced to move into the asymptotic phase too early.  This is a place
where we enter the second step, in which we expect occasional
``small'' deviations from the dominant behavior $\sim t_*^{-2}$, a
crucial pattern to maintain the decaying cosmological constant scenario.

\begin{figure}[h]
\hspace*{3.5cm}
\epsfxsize=10cm
\epsffile{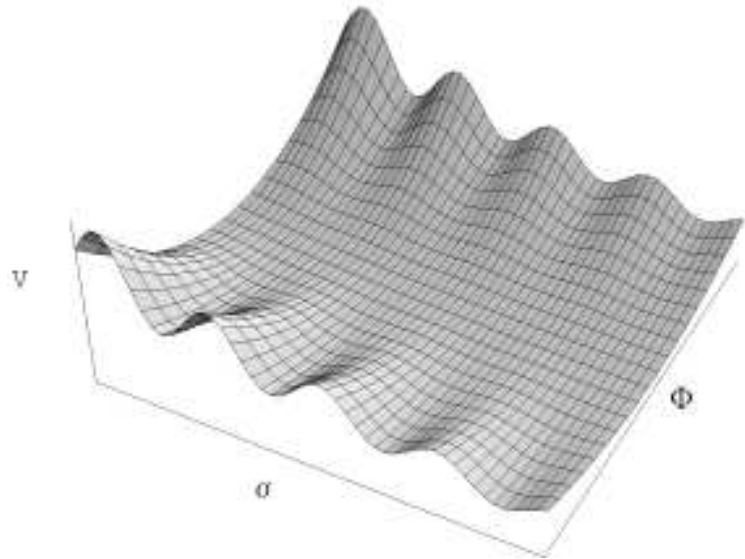}
\caption{The potential $V(\sigma, \Phi)$ given by \protect\reflef{ts-4}).  Along the central valley with $\Phi =0$, the potential reduces to the simpler behavior $\Lambda e^{-4\zeta\sigma}$ as given by \protect\reflef{ts-1}), but with $\Phi \neq 0$, it shows an oscillation in the $\sigma$ direction.}
\label{pot2}
\end{figure}

After a desperate search for the success, we decided to introduce another scalar field $\Phi$ that couples to $\sigma$ through the potential
\beq
V(\sigma, \Phi)=e^{-4\zeta\sigma}\tilde{V}(\sigma, \Phi),\quad\mbox{with}\quad
\tilde{V}= \Lambda+\half m^2\Phi^2\left[ 1+\gamma\sin(\kappa\sigma ) \right],
\label{ts-4}
\eeq
with $m, \gamma$ and $\kappa$  constants [\cite{plapp}], as also shown
graphically in Fig. \ref{pot2}.  This is supposed to be a potential in
E frame.  For $\Phi =0$, we go back to the exponential potential
\reflef{ts-1}) derived from $\Lambda$ simply by a conformal
transformation.  At this moment, we have no known theory at a more
fundamental level from which $\Phi$ and $V(\sigma,\Phi)$ derive, nor
we claim that \reflef{ts-4}) is a final result.  We nevertheless
emphasize that we then find solutions which are qualitatively
different from any of the results known so far, expecting to open up a new
perspective on the nature of the cosmological
constant problem.

In Section 1 we describe the model very briefly.  We then show a typical example of the solutions in Section 2, comparing the results with observations, before we go into some detailed discussion of the underlying mechanism in Section 3.  Final Section 4 is devoted to other related discussions.

\section{The model}

We reproduce the basic field equations in Refs. [\cite{plapp}].  In terms of $\tau =\ln t_*$ and $b =\ln a_*$ with $a_*$ the scale factor in E frame, the cosmological equations in spatially flat Robertson-Walker universe can be written as
\beqa
&&3 b'\:^2 = t_*^2 \left( \rho_s +\rho_{*m} \right), \quad\mbox{with}\quad \rho_s =\half \dot{\sigma}^2 +\half \dot{\Phi}^2 +V(\sigma,\Phi),
\label{ts-5} \\
&&\sigma'' +(3 b' -1)\sigma' + t_*^2 e^{-4\zeta\sigma}\left( -4\zeta\tilde{V} +\frac{\partial \tilde{V}}{\partial\sigma}  \right)= 0, \label{ts-6} \\
&&\Phi'' +(3 b' -1)\Phi' + t_*^2 e^{-4\zeta\sigma}\frac{\partial \tilde{V}}{\partial\Phi} = 0,  \label{ts-7} \\
&&\rho_{*m}' +4b'\rho_{*m}= 0, \label{ts-8} 
\eeqa
where a prime implies a derivative with respect to $\tau$.  We seek solutions of a spatially uniform $\sigma$.  We also assumed the radiation-dominated universe, for simplicity at this moment.  Due to the relation $H_* = (da_*/dt_*)/a_* =t_*^{-1}b'$, we find that choosing $\tau$ as an independent time variable has an advantage that the coefficients of the frictional force in \reflef{ts-6}) and \reflef{ts-7}) are now constant as far as $a_*(t_*)$ expands according to a power-law.  On the other hand, the potential in these equations is multiplied by $t_*^2 =e^{2\tau}$.  We use the reduced Planckian unit system.\footnote{Units of length, time and energy are $8.09 \times 10^{-33}{\rm cm}, 2.70 \times 10^{-43}{\rm s}$ and $2.44\times 10^{18}{\rm GeV}$, respectively. Note that the present age of the universe $\sim 10^{10}{\rm y}$ corresponds to $\sim 10^{60}$.}

We choose the initial time of integration to be $10^{10}$, somewhat later than the time when the reheating process is supposed be completed, though the true initial conditions for the classical fields should be given at a much earlier time.

\section{An example}

We give an example of the solution of \reflef{ts-5})-\reflef{ts-8}),
as shown in Fig. \ref{fx1}.  We came across this solution by a somewhat random search for the parameters and the initial values.  We have not attempted to scan the whole parameter space, though the values have been constrained essentially of the order one in the reduced Planckian unit system.

At the same time, we fine-tuned the parameters and initial values {\em moderately} in order to obtain considerable amount of $\Omega_{\Lambda}$  at the age around 10 Gy, where  we  find a crossing between $\rho_s$ and $\rho_{*m}$.  This naturally induces  an extra acceleration of the scale factor $a_*$, a ``mini-inflation."  In Fig. \ref{fx2}, a magnified view of the lower diagram of Fig. \ref{fx1}, we present more detailed behaviors of the densities, from which we can calculate  $\Omega_{\Lambda}$ and $h$, the Hubble constant in units of 100 km/sec/Mpc, for several values of the assumed present age of the universe, as listed in the ``first set" of Table \ref{rm3t1}.

It may appear that $\Omega_{\Lambda}$ and $h$ shown in the first set are shifted slightly to the higher side of those of the recent data; $\Omega_{\Lambda}\sim 0.7, h= 0.65, t_{*0}=1.3\times 10^{10}{\rm y}$ as summarized in [\cite{free}].  Instead of searching for another set of parameters, we exploit the fact that the eqs. \reflef{ts-5})-(\ref{ts-8}) are invariant under the rescaling $t_* \rightarrow \lambda t_*$ combined with $\Lambda \rightarrow \lambda^2\Lambda, m \rightarrow \lambda m, \rho_{*m,s}\rightarrow \lambda^2 \rho_{*m,s}$.  By choosing $\lambda = 0.7762$ to shift $\log t_* =60.11$ to 60.00, for example, we obtain the ``second set" of Table \ref{rm3t1}, yielding somewhat lower values of $\Omega_{\Lambda}$ and $h$.  It seems likely that we are able  to adjust ourselves
to almost any of the observational results which will be updated in the future with less uncertainties.

\begin{table}[tbh]
\bcent
\begin{tabular}{||ll||ll | ll||} \hline
\raisebox{-.8em}{$t_{*0}(10^{10}{\rm y})$} & \raisebox{-.8em}{$\log t_{*0}$}  
&\multicolumn{2}{c|}{\raisebox{-0.3em}{First set}} &\multicolumn{2}{c||}{\raisebox{-0.3em}{Second set}} \\
& &\ \ $\Omega_{\Lambda}$&\ \ $h$&\ \ $\Omega_{\Lambda}$&\ \ $h$  \\ \hline
1.1 & 60.11 &\ \ 0.62 & \ \ 0.81 &\ \ 0.46& \ \ 0.73 \\
1.2 & 60.15 &\ \ 0.67 & \ \ 0.77 &\ \ 0.52& \ \ 0.69 \\ 
1.3 & 60.18 &\ \ 0.72 & \ \ 0.74 &\ \ 0.56& \ \ 0.65 \\ 
1.4 & 60.22 &\ \ 0.76 & \ \ 0.72 &\ \ 0.62& \ \ 0.63 \\ \hline
\end{tabular}
\ecent
\caption{The assumed ages $t_{*0}, \log t_{*0}$ in the Planckian unit system, $\Omega_{\Lambda}$ and $h$ for the solution in Figs. \protect\ref{fx1} and \protect\ref{fx2} (First set), and for the rescaled age and other parameters (Second set).} 
\label{rm3t1}
\end{table}

We list other marked features shown in Fig. \ref{fx1}.  (i) We have
two plateau periods and the associated two mini-inflations.  We have
other examples with even more mini-inflations (with none as well).
This would imply that what we are witnessing at the present time may
not be a once-for-all event, but can be only one of the repeated
phenomena in the whole history of the universe.  In this way we can make the coincidence problem less severe. (ii)
The mini-inflation which has just started at the present time will
not last forever.  (iii)  In lower diagram of Fig. \ref{fx1}, we find that around the time of nucleosynthesis ($\log t_* \sim 45$), $\rho_s$ stays several orders smaller than $\rho_{*m}$.  Corresponding to this, the middle diagram of Fig. \ref{fx1} shows that the effective exponent $\alpha_*\equiv t_*H_*=b'$ stays at 0.5. It thus follows that  the conventional analysis of nucleosynthesis is left unaffected by the presence of the scalar fields.  This is a crucial criterion by which we can select acceptable solutions.  (iv) Around the present epoch, our solution does not behave like a scaling or a tracker solution [\cite{qta}]. Our solutions correspond exactly to the analyses in Refs. [\cite{garna},\cite{perlm}] including simply a cosmological constant.  The proposed tests on the equation of state $p/\rho$ [\cite{qtb}] might be interpreted as selecting models of the scalar field(s) rather than distinguishing them from the models of a purely constant $\Lambda$.  (v) A sufficiently long plateau of $\sigma$ may explain why the time-variability of some coupling constants as suspected from the time evolution of $\sigma$ have been constrained to the level much below $t_{*0}^{-1}\sim 10^{-10}{\rm y}^{-1}$ [\cite{tv},\cite{fon}].

\section{The mechanism}

Let us recall that already in the system of $\sigma$ alone there is a tendency that $\sigma$ is likely overdamped, because once it begins to move toward infinity, the exponential potential decreases so fast that it is only decelerated by the cosmological frictional force eventually to a standstill on the middle of the potential slope.   With $\Phi\neq 0$, on the other hand, there is now a chance that $\sigma$ is trapped in one of the minima of the sinusoidal potential.  With a still persistent frictional force, $\Phi$ will stay also perching on the slope.  This will continue to provide the potential wall of $\sigma$, thus enhancing the chance for $\sigma$ to stay sufficiently long, giving a sufficiently large $\Omega_{\Lambda}$.

In the mean time, the strength of the potential in \reflef{ts-6}) will grow due to the factor $t_*^2$.  The potential wall gets increasingly steeper, helping the energy of $\sigma$ to build up.  For the same reason the force acting on $\Phi$ grows also.  The exponent $f(\tau,\sigma)\equiv 2\tau -4\zeta\sigma$, which has been negative for $\sigma$ which had advanced sufficiently, will increase again.  If $f(\tau,\sigma) \sim 0$ is reached, $\Phi$ is pushed downward for the central valley $\Phi=0$.  Consequently, the potential wall which has been confining $\sigma$ finally collapses, releasing the accumulated energy to ``catapult" $\sigma$ forward, as shown in Fig. \ref{fx3}, a magnified plot of the upper diagram of Fig. \ref{fx1}.

We may expect that $\sigma$ is again decelerated and trapped by a potential minimum, which has in fact quite wide a basin of attraction, hence repeating nearly the same pattern as before, as shown in Fig. \ref{fx1}.  It may also happen, on the other hand, that $\sigma$ fails to be trapped.  Then without a potential wall for forced confinement, $\sigma$ will goes smoothly into the asymptotic stage as in the model with a single scalar field.  Figs. \ref{fy1}--\ref{fy3} provide an example of the behavior of this type. We find no crossing between the two densities, thus with $\Omega_{\Lambda}$ reaching only to 0.29, hence without extra acceleration of the scale factor around the present epoch.

Since the probability of being trapped is limited, however, the system will be, perhaps after some number of repeated mini-inflations, ultimately in the asymptotic behavior for {\em any} initial values.

\section{Discussions}

As we admitted, the potential \reflef{ts-4}) discovered on a
try-and-error basis is still tentative.  A crucial ingredient is obviously the presence of a sufficient number of minima with respect to
$\sigma$, each with a considerable basin of attraction.  These traps
should also depend on another dynamical field.  Other more general potentials
might be found resulting in the unique feature that mini-inflations can 
occur repeatedly, in connection with the interlacing pattern of $\rho_s$ and $\rho_{*m}$.

There are several other points to be discussed.  First, there is an issue on what CF is a physical CF.  Our calculations have been carried out in E frame.  In this connection we point out that in the prototype BD model the scalar field is assumed to be decoupled from the matter part of the Lagrangian in J frame in order to maintain Weak Equivalence Principle (WEP).  It then follows that masses of matter particles are time-independent even if the scalar field evolves.  In this sense J frame should be a physical CF, because we analyze nucleosynthesis, for example, on the basis of quantum mechanics with particle masses taken as constant.  We find, however, that the prototype BD model with $\Lambda$ included entails an asymptotically {\em static} (radiation-dominated) universe, to which the universe approaches oscillating or contracting [\cite{yf3}].   Although this conclusion depends on the simplest choice of the non-minimal coupling term as well as a purely constant $\Lambda$, we must modify the model in a rather contrived manner.\footnote{If we multiply $\Lambda$ by $\phi^{\ell}\ (\ell \neq 2)$, the scale factor $a(t)$ in J frame is found to behave like $a(t)=t^{\alpha}$ with $\alpha =(1/2)(\ell /(\ell -2))$.}

\begin{figure}[h]
\hspace*{2.5cm}
\epsfxsize=11cm
\epsffile{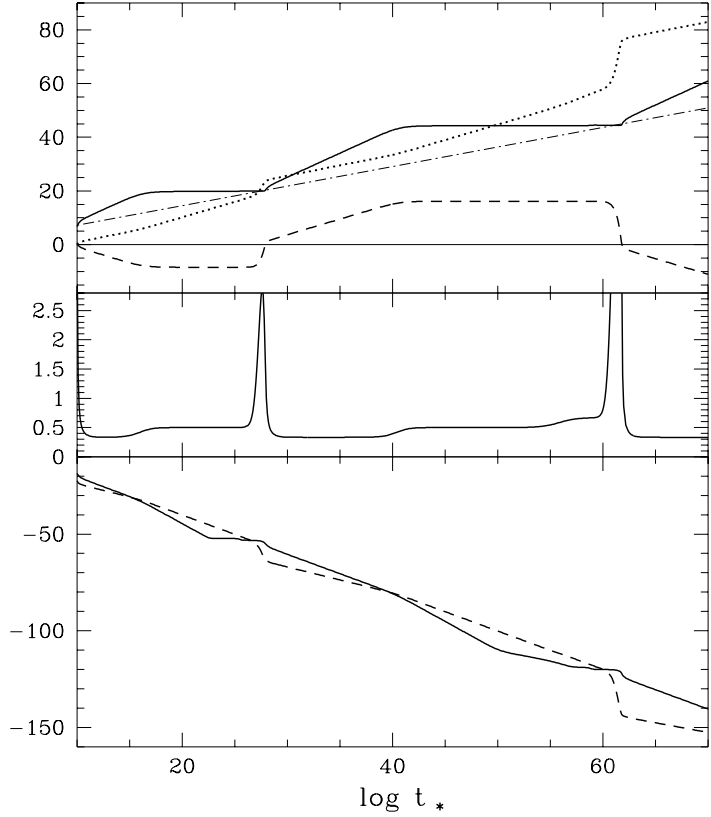}
\caption{An example of the solution.  Upper diagram: $b=\ln a_*$
(dotted), $\sigma$ (solid) and $2 \Phi$ (dashed) are plotted against
$\log t_*$.  The present epoch corresponds to $\log t_* =60.1-60.2$,
while the primordial nucleosynthesis must have taken place at $\log t_* 
\sim 45$.   The parameters are $\Lambda =1, \zeta = 1.5823, m= 4.75, 
\gamma = 0.8, \kappa = 10, \zeta_{\rm d} = 0.005$ (See Sect. 5 for the
effect of the dust component).  The initial values at $t_1 = 10^{10}$
are $\sigma _1=6.7544, \sigma'_1 =0$ (a prime implies differentiation
with respect to $\tau=\ln t_*$),$ \Phi_1 = 0.21, \Phi'_1 = -0.002,
\rho_{1 \rm rad}= 3.7352 \times 10^{-23}, \rho_{1 \rm dust}=4.0 \times
10^{-45}$.  The dashed-dotted straight line represents the asymptote of $\sigma$ given by $\tau /(2\zeta)$.  Notice long plateaus of $\sigma$ and
$\Phi$, and their rapid changes  during relatively ``short''
periods.   Middle diagram: $\alpha_* =b' = t_* H_*$ for an effective
exponent in the local power-law expansion $a_*\sim t_*^{\alpha_*}$ of
the universe.  Notable leveling-offs can be seen at 0.333, 0.5 and 0.667
 corresponding to the epochs dominated by  the kinetic terms of the
scalar fields, the radiation matter and the dust matter, respectively.
Lower diagram: $ \log\rho_s$ (solid), the total energy density of the
$\sigma-\Phi$ system, and $\log\rho_{*m}$ (dashed), the matter energy density.
  Notice an interlacing pattern of $\rho_s$ and $\rho_{*m}$, still obeying 
$\sim t_*^{-2}$ as an overall behavior.  Nearly flat plateaus of
$\rho_s$ precede before $\rho_s$ overtakes $\rho_{*m}$, hence with
$\Omega_{\Lambda}$ passing through 0.5.
}
\label{fx1}
\end{figure}

\begin{figure}[h]
\hspace*{2.5cm}
\epsfxsize=11cm
\epsffile{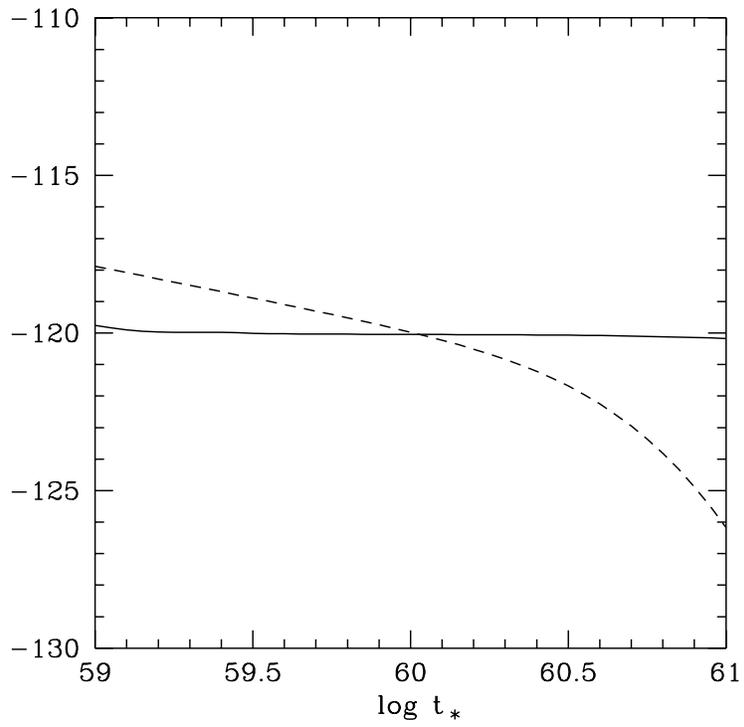}
\caption{Magnified view of $\log\rho_s$ (solid) and $\log\rho_{*m}$ (dashed) in the lower diagram
of Fig. \protect\ref{fx1} around the present epoch.  Note that
$\rho_s$ is very flat in this diagram extending back to the past
of $z=5.2-6.9$ for the assumed age $(1.1-1.4)\times 10^{10}{\rm y}$.
} 
\label{fx2}
\end{figure}

\begin{figure}[h]
\hspace*{2.5cm}
\epsfxsize=11cm
\epsffile{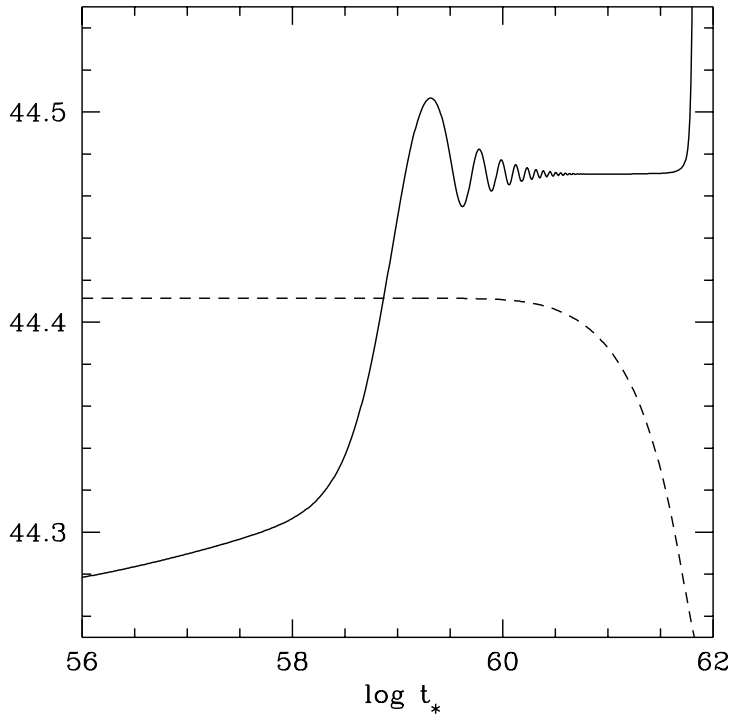}
\caption{Magnified view of $\sigma$ (solid) and $0.02 \Phi+44.25$ (dashed) in the upper diagram of
Fig. \protect\ref{fx1}.  Note that the vertical scale has been expanded by approximately 330 times as
large compared with Fig. \protect\ref{fx1}.  
With the time variable $\tau = \ln t_*$, the potential $V$ grows as
the multiplying factor $t_*^2=e^{2\tau}$.  The potential wall for
$\sigma$ becomes increasingly steeper, thus confining $\sigma$ further
to the bottom of the potential, as noticed by an oscillatory behavior of $\sigma$ with ever increasing frequency measured in $\tau$.  (In this particular example, $\sigma$ was trapped finally long after it had been decelerated to settle near one of the maxima of the potential.)  The growing $V$ causes $\Phi$ eventually to fall downward, resulting in the collapse of the confining potential wall.  The stored energy is then released to unleash $\sigma$.}
\label{fx3}
\end{figure}

As another approach we proposed to modify the mass terms of matter
fields guided by the invariance under global scale transformation
(dilatation) [\cite{yf3}], making E frame an approximately physical
CF.  WEP is then violated but can be tolerated within the constraints
from the fifth-force-type experiments.  Leaving the details in
Ref. [\cite{yf3}], however, we emphasize that we can evade large part of the complications in the choice of CF if we choose the type of solutions as exemplified in Fig. \ref{fx1}, because $\sigma$ stays constant over the period which covers nucleosynthesis and the present time.  Not only $\sigma$ and hence particle masses at the time of nucleosynthesis remain the same as today, but also a constant scalar field makes any difference among different CFs trivial.

Strictly speaking, however, we must include dust matter to solve the
equations beyond the time of recombination, $t_{* \rm eq}\sim
10^{55}$.  For this purpose we in fact added $\zeta_{\rm d}t_*^2
\rho_{*m}$ and $-\zeta_{\rm d}t_* \sigma'\rho_{*m}$ to the right-hand
sides of \reflef{ts-6}) and \reflef{ts-8}), respectively,
simultaneously replacing $4b'$ on the left-hand side of \reflef{ts-8})
by $3b'$.  We used the coupling strength estimated tentatively as
$\zeta_{\rm d}= 0.005$ determined as $\sim 1/4$ of the coefficient
$c_N \sim 0.02$ in eq. (58) of Ref. [\cite{yf3}].  Effects of these
modifications  turned out rather minor, though we recognize a slight distortion 
of  the otherwise flat $\rho_s$ for $50\lsim \log t_* \lsim
57$ in the lower diagram of Fig. \ref{fx1}.  We consider this as the maximally expected effect.

\begin{figure}[h]
\hspace*{2.5cm}
\epsfxsize=11cm
\epsffile{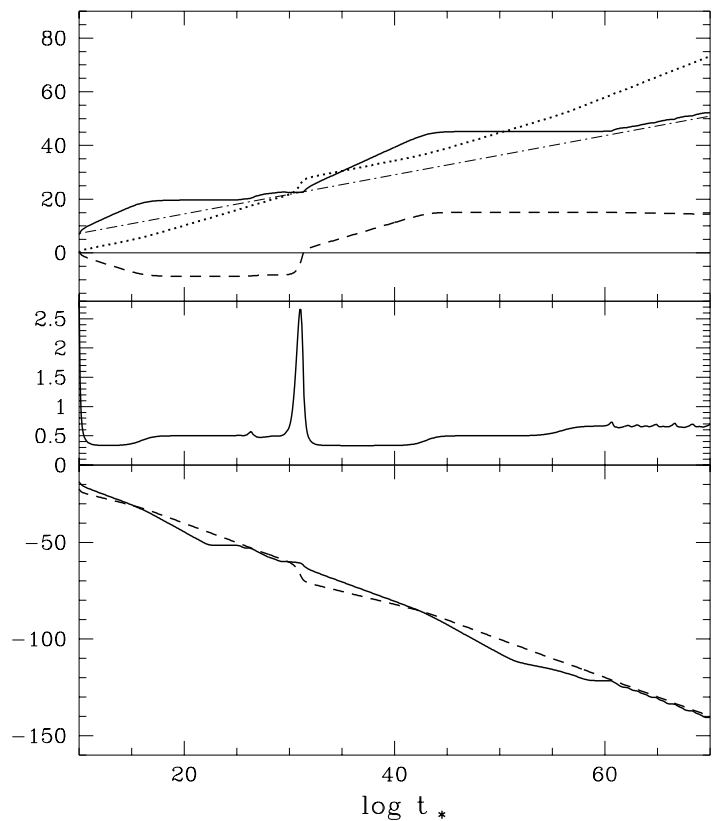}
\caption{An example of the solution without mini-inflation around the present epoch, to be compared with Fig. \protect\ref{fx1}.  The qualitative change of the behavior is a consequence of choosing a slightly different value $\sigma _1=6.7764$ as compared with 6.7544 in Fig. \protect\ref{fx1}, with the rest of parameters remaining the same.  In the upper diagram we find $\sigma$ making a smooth transition to the asymptotic behavior without being catapulted. Also  $\Phi$ approaches the value zero so slowly that it does not look it does in the limited range of time.   }
\label{fy1}
\end{figure}

\begin{figure}[h]
\hspace*{2.5cm}
\epsfxsize=11cm
\epsffile{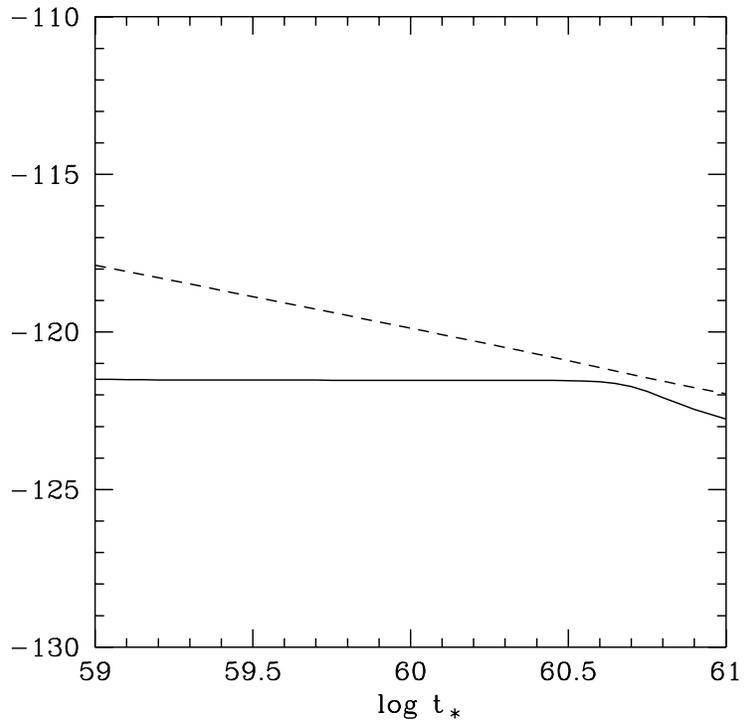}
\caption{Magnified view of $\log\rho_s$ (solid) and $\log\rho_{*m}$ (dashed) in the lower diagram
of Fig. \protect\ref{fy1} around the present epoch, to be compared with Fig. \protect\ref{fx2}.  No crossing is
seen, with $\Omega_{\Lambda}$ reaching only 0.29.} 
\label{fy2}
\end{figure}

\begin{figure}[h]
\hspace*{2.5cm}
\epsfxsize=11cm
\epsffile{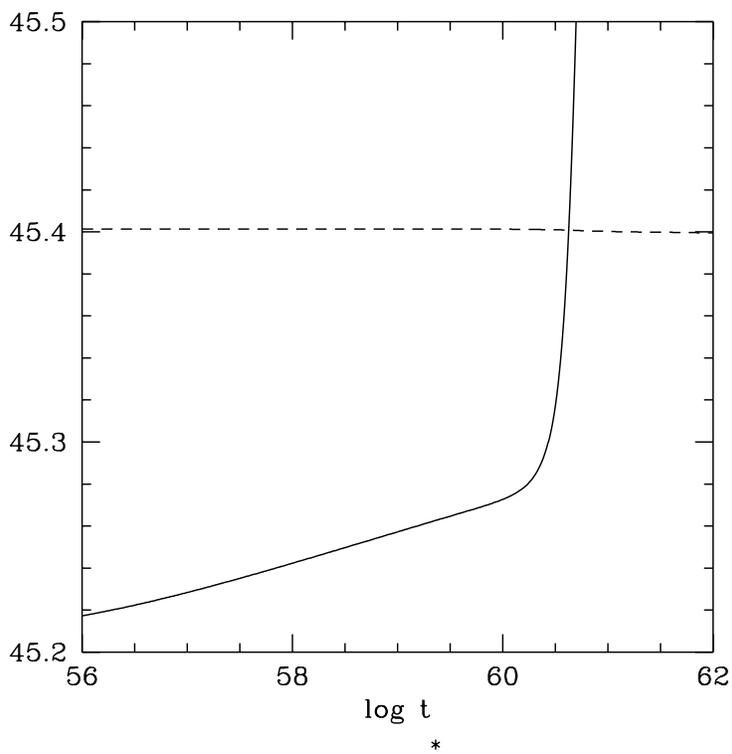}
\caption{Magnified view of $\sigma$ (solid) and $0.02 \Phi+45.25$ (dashed) in the upper diagram of
Fig. \protect\ref{fy1}.  Note that no oscillatory behavior is seen for $\sigma$  plotted in the same rate of magnification as in
Fig. \protect\ref{fx3}.  This is due to the absence of the force which
has protected $\sigma$ from flowing out as the potential grows as $t_*^2$. }
\label{fy3}
\end{figure}

As another aspect of our solution, we should mention the sensitivity on initial values. The absence of the second mini-inflation in Fig. \ref{fy1} is a consequence of a slight change of the initial value $\sigma_1$ from 6.7544 to 6.7744 chosen in Fig. \ref{fx1}.  We have in fact the solutions without any mini-inflation for $\sigma_1$ in between.  According to our experience,   we come across a solution of two mini-inflations rather easily if we scan $\sigma_1$ over the range $\Delta \sigma_1 \sim 0.6$ (corresponding to the change $2\pi$ of $\kappa\sigma_1$ with a pitch 0.01, for example.  More details will be reported elsewhere.

Traditionally we have made it a goal to make predictions depending  as little as possible on as few as possible number of initial conditions.  Our solutions do not seem to meet this condition.  We have at least four initial values for $\sigma$ and $\Phi$.  Moreover, the result depends on them very sensitively.   One may suspect chaotic behaviors in nonlinear dynamics of more than three degrees of freedom; we have at least four of them from the two scalar fields.

As it turns out, however, the final destination of the solutions in phase space is a fixed-point attractor instead of any kind of strange attractors, as long as the cosmological frictional force continues to be effective.  Our solutions show only  chaos-{\em like} behaviors rather than authentic chaotic behaviors as  discussed in Refs. [\cite{corn}].  A conventional deterministic view  may not be maintained.    In contrast to the long-held belief in physics, it seems that we are forced to accept a new natural attitude if the nonlinearity of the equations is indeed crucial to understand the issue of the cosmological constant [\cite{mor}].

Finally we point out that the exponential potential in E frame always emerges even if $\Lambda$ in J frame is multiplied by a monomial $\phi^{\ell}\ (\ell\neq 4)$, whereas a power-law potential $\sigma^{\ell}$ in E frame, as discussed in many references, can be interpreted as coming from a somewhat awkward choice of the J frame potential $\phi^4 (\ln \phi)^{\ell}$ added to the prototype BD model. 

\bigskip
\begin{center}
{\large\bf Acknowledgments}
\end{center}

I thank Naoshi Sugiyama, Akira Tomimatsu and Kei-ichi Maeda for many valuable discussions. I am also grateful to Ephraim Fischbach for a useful suggestion on part of the draft.
\bigskip

\begin{center}
{\large\bf References}
\end{center}
\begin{enumerate}
\item\label{plapp}Y. Fujii and T. Nishioka, Phys. Lett. {\bf B254}, 347(1991); Y. Fujii, Astropart Phys. {\bf 5}, 133(1996).
\item\label{garna}P. Garnavich {\it et al.}, Astrophys J., {\bf 493}, L3(1998).
\item\label{perlm}S. Perlmutter {\it et al.}, Nature {\bf 391}, 51(1998); Astrophys J., astro-ph/9812133.
\item\label{free}W.L. Freedman, astro-ph/9905222.
\item\label{yf3}Y. Fujii, Prog. Theor. Phys. {\bf 99}, 599(1998).
\item\label{yf1}A.D. Dolgov, {\sl The very early universe}, Proceedings of Nuffield Workshop, 1982, eds. B.W. Gibbons and S.T. Siklos, Cambridge University Press, 1982; Y. Fujii and T. Nishioka, Phys. Rev. {\bf D42}, 361(1990).
\item\label{qta}P.J.E. Peeble and B. Ratra, Astrophys J., {\bf 325}, L17(1988); B. Ratra and P.J.E. Peeble, Phys. Rev. {\bf D37}, 3406(1988); I. Zlatev, L. Wang and P.J. Steinhardt, Phys. Rev. Lett. {\bf 82}, 896(1999); A.R. Liddle and R.J. Scherrer, Phys. Rev. {\bf D59}, 023509(1999).
\item\label{qtb}T. Chiba, N. Sugiyama and T. Nakamura, MNRAS {\bf 289}, L5(1997); R.R. Caldwell, R. Dave and P.J. Steinhardt, Phys. Rev. Lett. {\bf 80}, 1582(1998); W. Hu, D.J. Eisenstein and M. Tegmark, Phys. Rev. {\bf D 59}, 023512(1998); G. Huey, L. Wang, R.R. Caldwell and P.J. Steinhardt, Phys.Rev. {\bf D59}, 063005(1999).
\item\label{qtc}P.G. Ferreria and M. Joyce, Phys. Rev. Lett. {\bf 79}, 4740(1997); Phys. Rev. {\bf D58}, 023503(1998); F. Rosati, hep-ph/9906427.
\item\label{qtd}J. Frieman and I. Waga, Phys. Rev. {\bf D57}, 4642(1998); S.M. Carroll, astro-ph/9806099; P.J. Steinhardt, L. Wang and I. Zlatev, Phys. Rev. {\bf 59}, 123504(1999); T. Chiba, gr-qc/9903094; M.C.Bento and O. Bertolami, gr-qc/9905075; F. Perrotta, C. Baccigalup and S. Matarrese, astro-ph/9906066; J. Garriaga, M. Livio and A. Vilenkin, astro-ph/9906210.

\item\label{fon}Y. Fujii, M. Omote and T. Nishioka, Prog. Theor. Phys. {\bf 92}, 521(1992).

\item\label{tv}F. Hoyle, {\sl Galaxies, Nuclei and Quasars}, Heinemann (London), 1965; F.J. Dyson, Phys. Rev. Lett. {\bf 19}, 1291(1967); P.C.W. Davies, J. Phys. {\bf A5}, 1296(1972); A.I. Shlyakhter, Nature, {\bf 264}, 340(1976); ATOMKI Report A/1 (1983), unpublished; R.W. Hellings, et al, Phys. Rev. Lett. {\bf 51}, 1609(1983); L.L. Cowie and A. Songalia, Astrophys J., {\bf 453}, 596(1995); A. Godone, et al, Phys. Rev. Lett. {\bf 71}, 2364(1993); T. Damour and F. Dyson, Nucl. Phys. {\bf B480}, 37(1996); J.K. Webb, V.V. Flambaum, C.W. Churchill, M.J. Drinkwater and J.D. Barrow, Phys. Rev. Lett. {\bf 82}, 884(1999); Y. Fujii, A. Iwamoto, T. Fukahori, T. Ohnuki, M. Nakagawa, H. Hidaka, Y. Oura and P. M\"{o}ller, hep-ph/9809549.
\item\label{corn}J.N. Cornish and J. Levin, Phys. Rev. {\bf  D53}, 3022(1996); R. Easther and K. Maeda, Class. Quant. Grav. {\bf 16}, 1637(1999).
\item\label{mor}Y. Fujii, Proceedings of the XXXIIIrd Rencontres de Moriond, {\sl Fundamental Parameters in Cosmology}, eds. J. Tran Thanh Van, Y. Giraud-Heraud, F. Bopuchet, T. Damour, Y. Melier, 93-95, Les Arcs, France, January 1998, Editions Frontieres, gr-qc/9806089.
\end{enumerate}



\newpage

\mbox{}\\[2em]

\end{document}